\documentstyle[12pt,a4wide]{article}
\input epsf
\newcommand{\be}{\begin{equation}}
\newcommand{\ee}{\end{equation}}
\font\mybb=msbm10 at 11pt

\def\bb#1{\hbox{\mybb#1}}

\def\R {\bb{R}}

\newcommand{\news}{\setcounter{equation}{0}}

\newcommand{\bx}{{\bf x}}
\newcommand{\by}{{\bf y}}
\newcommand{\bnab}{\mbox{\boldmath $\nabla$}}

\def\ben{\begin{equation}}
\def\een{\end{equation}}
\def\bea{\begin{eqnarray}}
\def\eea{\end{eqnarray}}
\input amssym.def
\input amssym.tex
\begin{document}

\title{\vskip -80pt
\begin{flushright}
{\small DAMTP-2002-137}\\[1cm]
\end{flushright}
\bf \Large \bf Polyhedral Scattering of Fundamental Monopoles\\[30pt]
\author{ Richard A. Battye$^{1}$, 
Gary W.  Gibbons$^{2}$,\\ Paulina Rychenkova$^{3}$
 and Paul M. Sutcliffe$^{4}$\\[10pt]
\\{\normalsize $^{1}$ {\sl Jodrell Bank Observatory, Macclesfield, Cheshire SK11 9DL U.K.}}
\\{\normalsize {\sl $\&$  Department of Physics and Astronomy,
Schuster Laboratory,}}
\\{\normalsize {\sl University of Manchester, Brunswick St,
 Manchester M13 9PL, U.K.}}
\\{\normalsize {\sl Email : rbattye@jb.man.ac.uk}}\\
\\{\normalsize $^{2}$ {\sl Department of Applied Mathematics and
Theoretical Physics,}}
\\{\normalsize {\sl Centre for Mathematical Sciences, University of Cambridge,}}
\\{\normalsize {\sl Wilberforce Road, Cambridge CB3 0WA, U.K.}}
\\{\normalsize {\sl Email : G.W.Gibbons@damtp.cam.ac.uk}}\\
\\{\normalsize $^{3}$  {\sl Charles River Ventures  ,}}
{\normalsize {\sl 1000 Winter St, Waltham,}}
{\normalsize {\sl MA 02451 USA.}}
\\{\normalsize{\sl Email : paulina@crv.com }}\\
\\{\normalsize $^{4}$  {\sl Institute of Mathematics,}}
{\normalsize {\sl University of Kent at Canterbury,}}\\
{\normalsize {\sl Canterbury, CT2 7NZ, U.K.}}\\
{\normalsize{\sl Email : P.M.Sutcliffe@ukc.ac.uk}}\\}}
\date{November 2002}
\maketitle

\begin{abstract}
The dynamics of $n$ slowly moving fundamental monopoles in the 
$SU(n+1)$ BPS Yang-Mills-Higgs theory can be approximated by
geodesic motion on the $4n$-dimensional hyperk\"ahler 
Lee-Weinberg-Yi manifold. In this paper we apply a variational
method to construct some scaling
geodesics on this manifold. These geodesics describe the scattering of $n$
monopoles which  lie on the vertices of a bouncing polyhedron;
the polyhedron contracts from infinity to a point, representing the spherically
symmetric $n$-monopole, and then expands back out to infinity. 
For different monopole masses the solutions generalize to form 
bouncing nested polyhedra. The relevance of these results to the dynamics
of well separated $SU(2)$ monopoles is also discussed.

\end{abstract}

\vfill \eject

\section{Introduction}\news
The dynamics of slowly moving BPS monopoles can be approximated by
geodesic motion on the moduli space of static solutions, with
the metric determined by the kinetic part of the Lagrangian \cite{Ma,Stu}.
For two centred $SU(2)$ monopoles the moduli space is the Atiyah-Hitchin
manifold and the simplest geodesic corresponds to the $90^\circ$ scattering
of two monopoles in a head-on collision \cite{AH}. Unfortunately for
more than two $SU(2)$ monopoles the moduli space metric is not known
explicitly, except in the region where all the monopoles are
well separated \cite{GM}. Despite this fact some geodesics are known
\cite{HMM,HS1,HS3,Su}. They are obtained by the imposition of appropriate
spatial symmetries to yield a one-dimensional manifold of static solutions,
which is then automaticaly a geodesic, since the fixed point set of
a group action is always a totally geodesic submanifold.

For BPS monopoles with gauge group $SU(n+1)$ and maximal symmetry
breaking there are $n$ topological charges and correspondingly 
$n$ types of fundamental monopole, each of which carries a single
unit of one of these charges \cite{We}. If there is precisely
one fundamental monopole of each type then the moduli space is $4n$-dimensional
and equiped with the hyperk\"ahler Lee-Weinberg-Yi metric \cite{LWY},
which is known explicitly. The explicit form of the metric allows
the possibility of computing some geodesics and hence $n$-monopole 
scattering processes for any value of $n.$
 In this paper we apply a variational
method to construct some scaling geodesics on this manifold.
The approach is to look for central configurations in which the time
dependence is only in the form of an overall scaling of the monopole
positions. The resulting algebraic equations can then be written
as the critical points of a certain energy function \cite{Ry},
which we minimize using numerical methods. As examples, we find minimal
energy configurations, and hence geodesics, for all $n\le 20.$
The symmetries of these configurations are analyzed and suggest the
existence of icosahedral minima at $n=32$ and $n=72,$ which are also
constructed. In all cases the associated geodesics
describe the scattering of $n$
monopoles which lie on the vertices of a bouncing polyhedron, in
the following sense.
The polyhedron first contracts from infinity to a point,
which in the moduli space represents the spherically
symmetric $n$-monopole. The evolution then reverses with the 
monopoles located on the vertices of the same polyhedron, but which
is now expanding back out to infinity. 

For different monopole masses the above solutions generalize to 
form bouncing nested polyhedra. Our solutions also provide geodesics of the 
Gibbons-Manton metric \cite{GM}, and hence describe the scattering 
of $SU(2)$ $n$-monopoles, valid in the region where the monopoles
are well-separated. This reveals a connection with some geodesics
obtained earlier using symmetry arguments.  

\section{Polyhedral Scattering}\news
The $4n$-dimensional hyperk\"ahler Lee-Weinberg-Yi manifold is a
$T^n$ bundle over a $3n$-dimensional base space. 
For $i=1,..,n$ let $\theta_i\in S^1$ be the fibre coordinates
and $\bx_i\in\R^3$ be local coordinates in the base, which may
be thought of as the positions in $\R^3$ of each of the $n$ monopoles.
 The purely kinetic
Lagrangian associated with the metric has the form
\be
L=g_{ij}\dot\bx_i\cdot\dot\bx_j+g_{ij}^{-1}
(\dot\theta_i+{\bf W}_{ik}\cdot\dot\bx_k)
(\dot\theta_j+{\bf W}_{jl}\cdot\dot\bx_l)\,,
\ee
where we have used the Einstein summation convention, though this
is not to be used in the rest of the paper unless explicitly stated.
The quantities appearing in the above are given by
\bea
g_{ii}&=&m_i+\sum_{j\ne i}\frac{1}{|\bx_i-\bx_j|}\label{lwy1}\,,\\
g_{ij}&=&\frac{-1}{|\bx_i-\bx_j|}\,,\quad i\ne j\\
{\bf W}_{ii}&=&\sum_{j\ne i}{\bf w}_{ij}\,,\\
{\bf W}_{ij}&=&-{\bf w}_{ij},\quad i\ne j\label{lwy4}
\eea
and ${\bf w}_{ij}$ is the value at $\bx_i$ of the Dirac potential due
to the monopole at $\bx_j$, that is
\be
\bnab_j\times{\bf w}_{ji}=\frac{\bx_i-\bx_j}{|\bx_i-\bx_j|^3}.
\ee
In the above we have scaled out the magnetic charge of a monopole
and the positive constants $m_i$ are related to the monopole masses.

The $T^n$ isometry of the metric yields the $n$ conserved charges
(here we use the summation convention once more)
\be
Q_i=g_{ij}^{-1}(\dot\theta_j+{\bf W}_{jk}\cdot\dot\bx_k)\,,
\ee
so that the fibre coordinates are non-dynamical degrees of freedom.
In this paper we shall be concerned with monopoles with no electric
charge, so we set $Q_i=0$, for all $i=1,..,n.$
In this case the Lagrangian describing the motion in the base space is simply
\be 
L=\sum_i\left(m_i+\sum_{j\ne i}\frac{1}{x_{ij}}\right)\dot\bx_i^2
-\sum_{i}\sum_{j\ne i}\frac{1}{x_{ij}}\dot\bx_i\cdot\dot\bx_j.
\label{lag}\ee
where we have defined $\bx_{ij}=\bx_i-\bx_j,$ and
$x_{ij}=|\bx_{ij}|.$
The geodesic equations which follow from (\ref{lag}) are 
\be
m_k\ddot\bx_k=\sum_{j\ne k}\left(
\frac{\ddot\bx_{jk}}{x_{jk}}+\frac{\bx_{jk}|\dot \bx_{jk}|^2}{2x_{jk}^3}
-\frac{\dot\bx_{jk}\dot x_{jk}}{x_{jk}^2}\right)\,,
\label{gengeo}
\ee

As suggested in \cite{Ry} we now look for time-dependent homothetic
solutions of these equations, that is, solutions of the form
$\bx_k(t)=\alpha(t)\by_k,$ with constant $\by_k.$
Clearly, such solutions describe monopoles
in a fixed configuration, but with the overall scale of the configuration
evolving dynamically. Substituting this ansatz into (\ref{gengeo})
yields the equations
\be
m_k\by_k+C\sum_{j\ne k}\frac{\by_{jk}}{y_{jk}}=0\,,
\label{geo}
\ee
where $C$ is defined to be the quantity
\be
C=\frac{\dot\alpha^2}{2\ddot\alpha\alpha^2}-\frac{1}{\alpha}.
\label{alphaeqn}
\ee
Obviously, for a non-trivial solution of (\ref{geo}) to exist
the quantity $C$ must be a constant, and it turns out that
only a positive constant produces a physically acceptable solution.
By a rescaling of the time variable we may, without loss
of generality, set $C=1.$ The two constants which appear in
the general solution of (\ref{alphaeqn}) may be absorbed by a linear
transformation of $t,$ and the solution we require is given implicitly by
\be
t=\sqrt{\alpha+\alpha^2}+\frac{1}{2}\log(1+2\alpha+2\sqrt{\alpha+\alpha^2})\,,
\label{scale}\ee
for $t\ge 0.$
It is clear from (\ref{scale}), that the scale $\alpha(t)$ 
is a monotonically increasing function of $t,$ which for small
$t$ has the expansion $\alpha=t^2/4+...$ and for large $t$ has the
asymptotic form $\alpha\sim t.$   

Note that equations (\ref{gengeo}) are invariant under time-reversal
$t\mapsto -t$ and also spatial inversion of all the points 
$\bx_k\mapsto -\bx_k.$ So far we have only addressed the second part
of the geodesic motion, when $t\ge 0$ and the monopoles are moving
away from each other, but the first part with $t<0$ is simply obtained
by time-reversal in equation (\ref{scale}), so that the monopoles 
approach the origin from spatial infinity. 
Thus the monopoles bounce back off each other, rather than passing through
each other, which would have been the result if the time inversion was
accompanied by the spatial inversion $\bx_k\mapsto -\bx_k.$ 
The fact that the first of these scenarios is the correct one
can be seen by studying the Lee-Weinberg-Yi manifold
in the neighbourhood of the origin $\bx_k=0,$ for all $k.$ 
Although the metric appears to be singular at the origin, this is
merely a coordinate singularity, and if new coordinates are chosen 
appropriately 
(these are essentially polar coordinates but with the radial
variables related to the monopole positions by $r_k=\sqrt{|\bx_k|}$)
the metric is seen to be
flat in these new coordinates.
The fact that the squares of the monopole positions are related
to the flat coordinates is the reason that fundamental 
monopoles of different types bounce back upon collision; had the
metric been flat around the origin in the coordinates $\bx_k$ then
the monopoles would have passed through each other.

The problem of finding scaling geodesics has now been reduced to the
algebraic problem of finding sets of $n$ points $\by_k,$ which satisfy
(\ref{geo}) with $C=1.$ Our method is to use a variational approach,
based on the fact that (\ref{geo}) are the equations for
critical points of the energy function
\be
E=\frac{1}{2}\sum_i m_i|\by_i|^2-\sum_{i}\sum_{j<i}y_{ij}.
\label{energy}
\ee
In this formulation the problem has obvious similarities with the
classical problem of finding central configurations 
\cite{BGS} (which arise
in a similar way when a time-dependent homothetic ansatz is used
in Newton's equations of motion for gravitating point particles)
or equivalently solutions of the One Component Plasma (OCP) model \cite{Bau}.
The OCP model describes point charges immersed in a uniform background
of charge with the opposite sign. Hence there are two competing forces,
the first is an attraction towards the origin, represented by exactly
the same expression as the first term in (\ref{energy}), and repulsion
between the points, which in the OCP case is described by the Coulomb
energy. The second term in (\ref{energy}) plays a similar role in our 
problem as the Coulomb energy does in the OCP model. The contribution
of this non-positive term produces two-body particle repulsions which
can balance the attractive central force, producing stable minimal
energy configurations with finite non-zero separations. 

In the remainder of this section we shall restrict to the case when all the 
monopole masses are equal. By rescaling the positions $\by_k$ by the
inverse of this common mass we obtain the situation in which all
monopole masses are equal to unity, so for the rest of this section
we set $m_i=1$ for all $i=1,...,n.$

Although any critical point of the energy (\ref{energy}) will provide
us with a geodesic on the Lee-Weinberg-Yi manifold we shall concentrate
only on local minima, since these are the easiest to find numerically,
and ignore any saddle point solutions.
Presumably Leech's symmetric configurations \cite{Le},
 which are sets of particles
on a sphere in equilibrium under any force law between pairs of particles,
will also yield critical points of (\ref{energy}) if the particles are
allowed to move off the sphere, but are required to maintain all symmetries
of the spherical configuration. Leech's configurations consist of
an infinite family of polygons and bipyramids and a finite family 
with Platonic symmetry. 

The numerical scheme employed
is a multi-start gradient flow algorithm with randomly distributed
initial conditions. The energy function (\ref{energy}) 
has the obvious $SO(3)$ invariance associated with a spatial rotation
of all $n$ points, and also reflection symmetries which change the
sign of any one of the three Cartesian components of all the points. 
Up to the action of these symmetry groups, all the minimal 
energy solutions we find are unique. 

The case $n=1$ is trivial; the minimal energy solution is a single
point at the origin, with $E=0,$ and hence the scaling solution is
time-independent, so no geodesic is obtained. For two points the
minimal energy is $E=-1$ which occurs if $\by_1=-\by_2=(0,0,1),$ or
any spatial rotation of this configuration.  In other words, the two
monopoles are at antipodal points on the unit sphere. The associated
geodesic describes the head-on collision of two monopoles, in which
the spherically symmetric 2-monopole is formed, after which the
monopoles bounce back off each other. This scattering process was
first described by Connell \cite{Con}, who discovered that the metric
on the centred moduli space of two different fundamental $SU(3)$
monopoles is Taub-NUT with a positive mass parameter.

In Table 1 we present, for $2\le n\le 20$, the energy $E$ of the minimizing
configuration, its symmetry group $G,$ the distance from the origin
of the closest point $r_{\rm{min}},$ and the distance from the origin
of the furthest point $r_{\rm{max}}.$

\begin{table}
\centering
\begin{tabular}{|r|r|c|r|r|}
\hline
$n$ & $E$\ \ \ \ \  & $G$ & $r_{\rm{min}}$ & ${r_{\rm{max}}}$\\
\hline
   2&         -1.0000 &  $D_{\infty h}$&    1.0000 &    1.0000\\
   3&         -4.5000 &  $D_{3h}$ &    1.7321 &    1.7321\\
   4&        -12.0000 &  $T_d$ &    2.4495 &    2.4495\\
   5&        -24.5916&  $D_{3h}$ &    3.1018 &    3.1592\\
   6&        -43.9706&  $O_h$ &    3.8284 &    3.8284\\
   7&        -71.0162&  $C_1$ &    4.4782 &    4.5635\\
   8&       -107.5011&  $D_{4d}$ &    5.1841 &    5.1841\\
   9&       -154.5499&  $D_{3h}$ &    5.8376 &    5.8718\\
  10&       -213.5297&  $D_{4d}$ &    6.5099 &    6.5412\\
  11&       -285.6593&  $C_{2v}$ &    7.1648 &    7.2435\\
  12&       -372.7470&  $Y_h$ &    7.8819 &    7.8819\\
  13&       -475.3419&  $C_{2v}$ &    8.5186 &    8.5980\\
  14&       -595.4323&  $D_{6d}$ &    9.2142 &    9.2749\\
  15&       -734.0923&  $D_3$ &    9.8771 &    9.9279\\
  16&       -892.7338&  $T$ &   10.5541 &   10.5925\\
  17&      -1072.6591&  $D_{5h}$ &   11.2308 &   11.2368\\
  18&      -1275.2163&  $D_{4d}$ &   11.8834 &   11.9107\\
  19&      -1501.5794&  $C_{2v}$ &   12.5491 &   12.5987\\
  20&      -1753.4547&  $D_{3h}$ &   13.2348 &   13.2518\\
\hline
\end{tabular}
\caption{For $2\le n\le 20$ we list the energy $E$ of the minimizing
configuration, its symmetry group $G,$ the distance from the origin
of the closest point $r_{\rm{min}},$ and the distance from the origin
of the furthest point $r_{\rm{max}}.$}
\end{table}

In the examples in Table 1 where $r_{\rm{min}}$ is equal to $r_{\rm{max}}$ to
the accuracy presented they are in fact precisely equal, indicating that
all the points lie on the surface of a sphere of radius $r_{\rm{min}}=r_{\rm{max}}.$
For all the other cases it can be seen that $r_{\rm{min}}$ and $r_{\rm{max}}$
are very close in value, showing that all $n$ points lie close to,
but not exactly on, a sphere. As we shall see later, this feature appears
to persist for arbitrarily large values of $n,$ which contrasts sharply
with traditional central configurations with a Coulomb interaction,
where this property exists only for $n<13$ and beyond this value
there are multiple shells \cite{BGS}. It is interesting to note
that a scale invariant geometric energy function exists which also
yields minimal energy configurations on a single shell for all
numbers of points \cite{AS}. It would be interesting to try and
classify the properties of interaction potentials which produce 
only a single shell.

\begin{figure} 
\begin{center}
\leavevmode
\ \vskip -0cm
\epsfxsize=15cm\epsffile{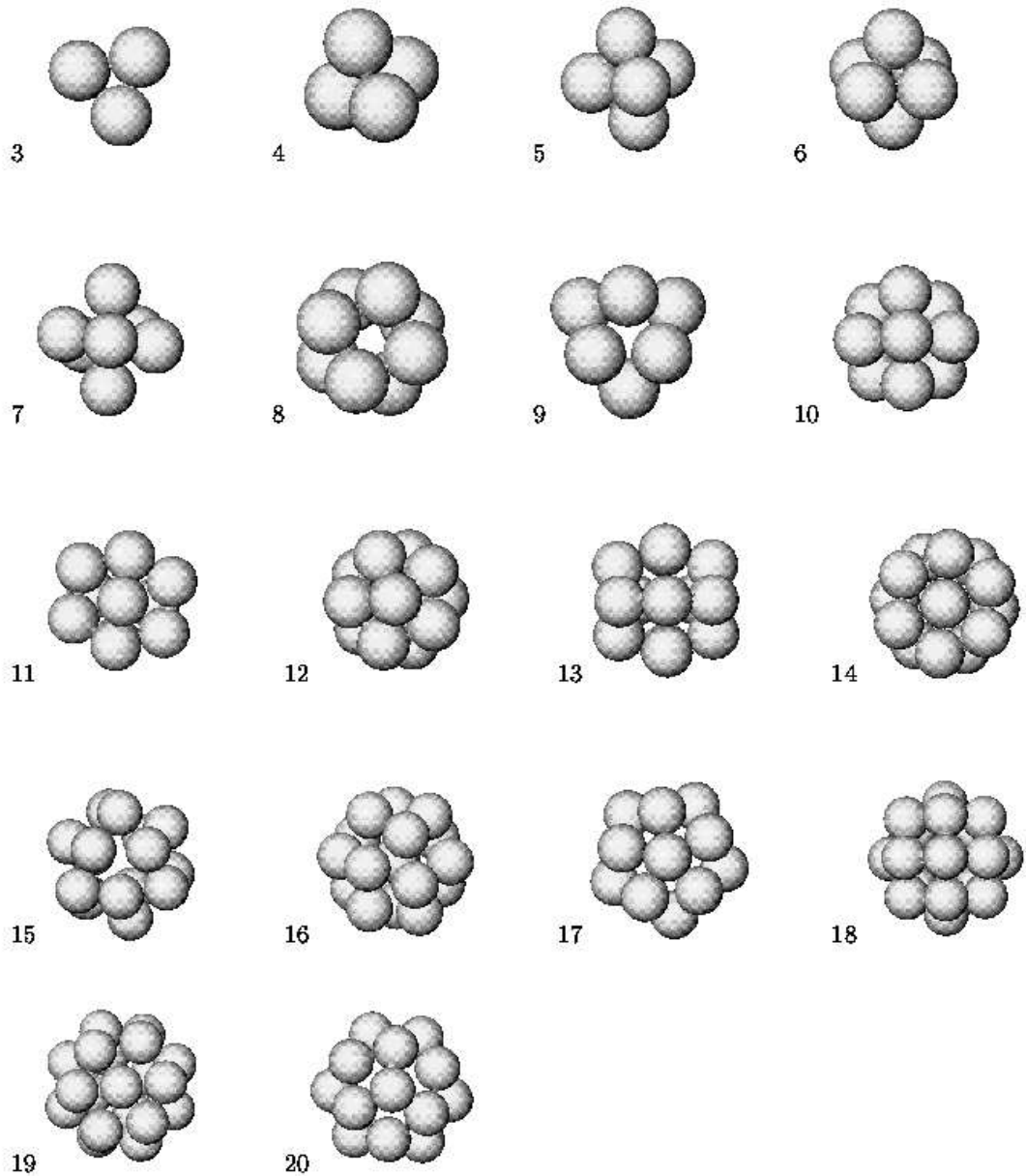}
\caption{For $3\le n \le 20$ we display the configurations of $n$ points
(not to scale) by plotting spheres around each of the points. In each case
the diameter of the spheres is equal to the minimal separation between
points, to emphasize the sphere packing behaviour.}
\label{balls}
\end{center}
\end{figure}

\begin{figure} 
\begin{center}
\leavevmode
\ \vskip -2cm
\epsfxsize=10cm\epsffile{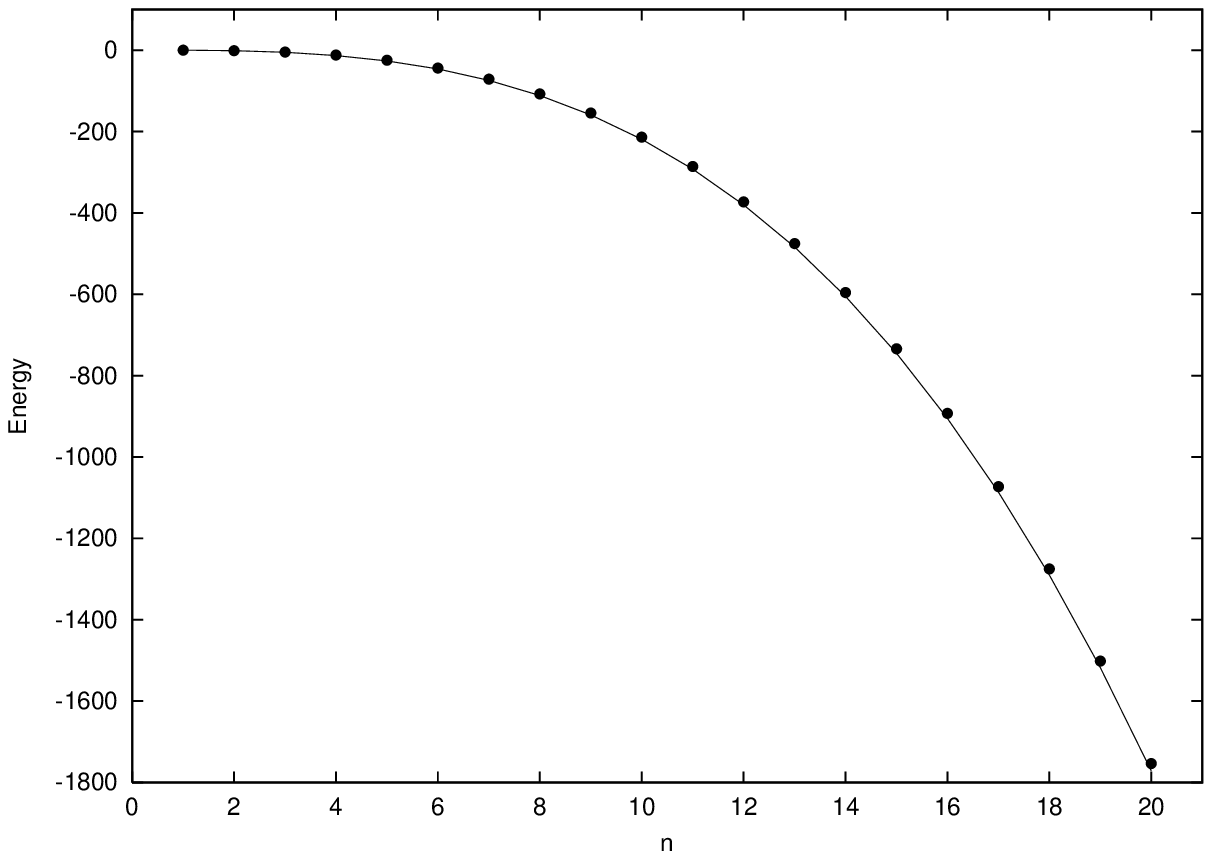}
\caption{The energy as a function of $n$ (circles) and the
estimate described in the text (curve).}
\label{enfit}
\end{center}
\end{figure}

In fig.~\ref{balls} we present, for $3\le n \le 20$,  the minimal
energy configurations of $n$ points by plotting spheres around each of
the points and in fig.~\ref{enfit} we plot their energy as a function of
$n$, for $n\le 20.$ In each case the diameter of the spheres is equal
to the minimal separation between points, to emphasize the sphere
packing behaviour. As seen from Table 1, the size of the configuration
(as measured by $r_{\rm{max}}$) grows with $n,$ so for clarity we do
not display the configurations to scale.  Three points lie on the
vertices of an equilateral triangle, with edge length equal to $3.$
For $n>3$ the points may be considered as forming the vertices of a
polyhedron, which generically is a deltahedron, that is, all faces are
triangular.  For example, four points lie on the vertices of a
tetrahedron with edge length equal to $4.$
For $n=4,6,8,12$ all points lie exactly on the surface of a sphere,
and in fact on the vertices of a tetrahedron,
octahedron, square antiprism and icosahedron, respectively.

As can be seen from Table 1 and fig.~\ref{balls} the points are
often arranged symmetrically, though the case
 $n=7$ is rather anomalous. There is an obvious
$D_{5h}$ symmetric candidate for the minimal
energy $n=7$ configuration, in which 5 points lie on the vertices of 
a regular pentagon and the two remaining points lie on the 5-fold symmetry
axis equidistant from the origin. This regular bipyramid is the obvious
generalization of the minimal $n=5$ configuration, which is a bipyramid
with a triangular base.
However, the minimal energy solution for $n=7$ is a symmetry breaking 
perturbation of the bipyramid. There are points at the north and south
poles of a sphere of radius 4.5635, and the remaining five points lie
in a roughly pentagonal distribution, but all with slightly different
heights above or below the equatorial plane and  different 
distances from the origin, which range from 4.4782 to 4.4825.
Clearly this prohibits any exact symmetry, even reflection symmetries,
so we label the symmetry group as $C_1,$ indicating no point symmetries.
As a check it is possible to minimize within the family of $D_{5h}$
symmetric configurations, yielding an energy of $E=-71.0156,$ which
is indeed slightly higher than the asymmetric minimum with $E=-71.0162.$

To gain insight into minimizing the energy function (\ref{energy})
(with $m_k=1$) it is useful to consider the restricted problem
in which all the points are constrained to lie on a sphere of
a given radius $\rho.$ The energy of this restricted problem is
given by
\be
E_\rho=\frac{1}{2}\rho^2n+\rho U\,,
\label{erho}
\ee
where
\be
U=-\sum_{i}\sum_{j<i}|{\bf Y_i}-{\bf Y_j}|\,,
\ee
for $n$ points ${\bf Y_k}={\bf y}_k/\rho$ restricted to lie on the
 surface of the unit sphere.
Minimization of the function $U$ for points on the unit sphere is a
problem in discrete geometry which was first posed almost fifty years
ago by Fejes T\'oth \cite{FT};
though it is usually phrased in terms of maximizing the sum of the 
mutual separations
$-U.$ There are a number of theorems proved about the extrema of this energy
function and in particular there is the lower bound \cite{Al}
\be
U > \frac{1}{2}-\frac{2}{3}n^2.
\ee
Using this result in (\ref{erho}) we obtain a lower bound for $E_\rho$
which we can then minimize over the radius $\rho,$
finding a minimum value at
\be
\rho=\frac{2n}{3}-\frac{1}{2n}\,,
\label{rho}
\ee 
to obtain the lower bound
\be
E_{\rm{sphere}}>-\left(\frac{2}{9}n^3-\frac{1}{3}n+\frac{1}{8}\right)\,,
\label{bound}
\ee
where $E_{\rm{sphere}}$ denotes the energy (\ref{energy}) (with $m_k=1$)
under the restriction that all points lie at the same distance from the
origin. Clearly when we drop this restriction we have no rigorous lower
bound for the unrestricted energy (\ref{energy}) which is our main concern,
but since our numerical results suggest that in all the minimal energy
configurations the points lie very close to the surface of a sphere then
we expect that the quantity in (\ref{bound}) will be a good estimate
of the minimal energy value, though it will tend to be slightly lower.
In fig.~\ref{enfit} we plot this estimate as the curved line. Clearly
the above expectations appear to be realized, in that the estimate is
close to the true value, but bounds it from below.
\begin{figure} 
\begin{center}
\leavevmode
\ \vskip -2cm
\epsfxsize=10cm\epsffile{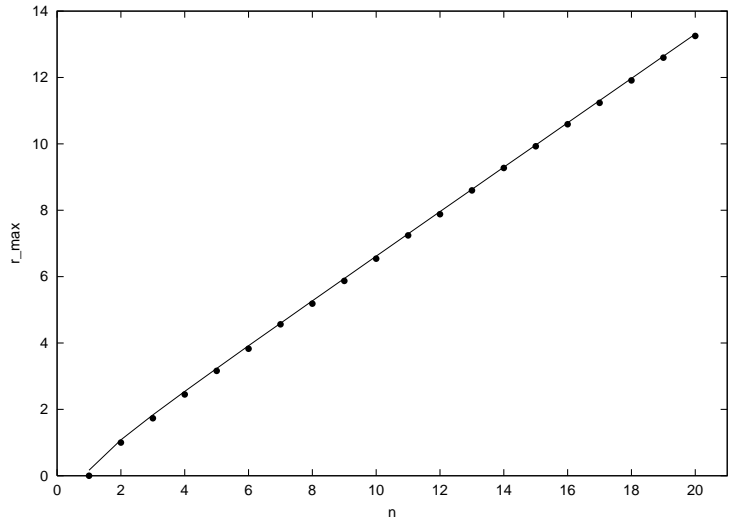}
\caption{The size of the configuration $r_{\rm{max}}$ as a function of $n$ 
(circles) and the estimate $\rho$ described in the text (curve).}
\label{size}
\end{center}
\end{figure}
In fig.~\ref{size} we compare the estimate (\ref{rho}) (curve) for the size
of the configuration with the numerical values as measured by $r_{\rm{max}}$
(circles). Again it can be seen that the estimate is quite accurate.

The above discussion suggests that our minimal energy solutions are
closely related to those which maximize the sum of the mutual
separations for points on a sphere. For $n=2,3,4,8,12$ it is proved
that the extremal configurations for this sphere problem are
the dipole, triangle, tetrahedron, octahedron and icosahedron
respectively (see \cite{An} and references therein), which agrees
with our results. For $n\le 10$ this problem was investigated
numerically by Berman and Hanes \cite{BH}
and more recently a comprehensive numerical investigation has
been performed by Rakhmanov, Saff and Zhou \cite{RSZ},
 who studied all $n\le 200$
and obtained the symmetry groups of the extremal configurations.
A comparison of the symmetry groups in Table 1 with those that appear
in \cite{RSZ} reveals that the groups agree
\footnote{In \cite{RSZ} the symmetry group for 18 points
should read $D_{4d}$ not $D_{4h}.$} for all cases except $n=7.$ 
In \cite{RSZ} the symmetry group for $n=7$ is given as $C_2$ and
in \cite{BH} the configuration is described as two almost
antipodal points with the remaining five points sprinkled around
an equatorial band. Our configuration for $n=7$ is therefore
consistent with a small deformation of the spherical extremal
solution, which itself has little symmetry.

For all $1\le n\le 20$ we found only one local minimum of the energy
for each value of $n$, except for $n=16.$ For $n=16$ the global minimum
with $E=-892.7338$ has tetrahedral symmetry $T,$ but we also found
a local minimum with energy $E=-892.7256$ and symmetry $D_{4h}.$
Once again this mirrors the situation in studying extremal problems
for points on a sphere \cite{EH}.

In order to further investigate the similarites between our solutions
and points on the sphere which maximize the sum of the mutual separations
we turn our attention to
configurations with icosahedral symmetry. 
For the sphere problem extremal configurations with icosahedral 
symmetry occur
for a sequence of points \cite{RSZ} which begins $n=12,32,72,..$.
As we have seen, for $n=12$ our solution has icosahedral symmetry,
with the points lying on the vertices of an icosahedron, so it is interesting
to compute the minimal energy solutions for $n=32$ and $n=72,$ to
see if they are icosahedrally symmetric.

\begin{figure} 
\begin{center}
\leavevmode
\ \vskip -0cm
\epsfxsize=10cm\epsffile{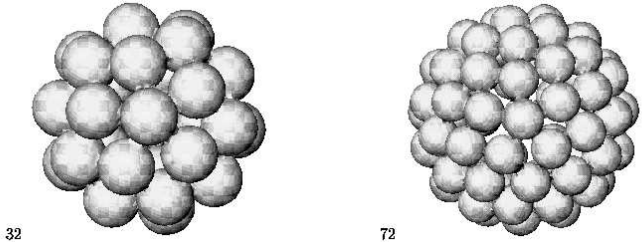}
\caption{For $n=32$ and $n=72$ we display the icosahedrally symmetric 
minimal energy configurations by drawing spheres around the monopoles.}
\label{icos}
\end{center}
\end{figure}

In fig.~\ref{icos} we display the minimal energy configurations for
$n=32$ and $n=72,$ using the same method as in fig.~\ref{balls}.
Both configurations contain a single shell and have icosahedral
symmetry, as predicted by analogy with the sphere problem.
For $n=32$ there are 12 points on the vertices of an icosahedron
at a distance of $21.2516$ from the origin and a further 20 points
at a distance of $21.2680$ from the origin. The associated polyhedron
is the dual of a truncated icosahedron and the energy is $E=-7233.0539.$
For $n=72$ the energy is $E=-82780.0335.$ There are 12 points at a distance
of $47.9338$ and 60 points at a distance of $47.9563.$

\section{Distinct masses and $SU(2)$ monopoles}\news

In this section we mention two extensions of the study we
have described so far.
The first is the obvious generalization to monopoles
which are not all of equal mass. As stated earlier, if all monopoles
have the same mass then varying this mass produces the same configuration
but scaled by the inverse common mass. Thus, the expectation for
sets of distinct masses is that multiple shells will arise,
 with monopoles grouped into shells according to their mass,
so that the heaviest monopoles sit closest to the origin, and with each
set of monopoles in a given shell being arranged on
the vertices of the polyhedron which arises in the minimization of the
relevant number of equal mass monopoles. This nested polyhedron picture
is consistent with the numerical results we have obtained. For example,
in the case of two distinct masses and four monopoles of each mass, we set
$m_i=1,\ $ for $i=1,..,4$ and $m_i=2,\ $ for $i=5,..,8$ the resulting
configuration is that the four heavy monopoles sit on the vertices of
 a tetrahedron at a distance 1.5940 from the origin and the four light
monopoles sit on the vertices of the dual tetrahedron scaled so
that they are at a distance 6.3543 from the origin. As another example,
with twelve light monopoles (with mass one) and six heavy monopoles
(with mass two) the nested polyhedra are an octahedron and an icosahedron
with scales 2.6954 and 13.8057 respectively, and oriented so as to 
preserve their common tetrahedral subgroups. Clearly the 
relative orientations of sets of nested polyhedra, together 
with their deformation as very distinct monopole masses are varied towards
equality are interesting problems which are likely to need substantial
investigation in each specific case.

The second extension we consider is to relate our results to
the dynamics of well separated $SU(2)$ monopoles.
Although, for $n>2$, the moduli space metric for $n$ $SU(2)$
 monopoles is not known explicitly, the asymptotic metric is known,
which is valid in the region where all the monopoles are well separated
\cite{GM}. This is the Gibbons-Manton metric and it is related to the
Lee-Weinberg-Yi metric through some sign changes. Explicitly, the
Gibbons-Manton metric is obtained by replacing equations 
(\ref{lwy1})-(\ref{lwy4}) by the equations
\bea
g_{ii}&=&m_i-\sum_{j\ne i}\frac{1}{|\bx_i-\bx_j|}\\
g_{ij}&=&\frac{1}{|\bx_i-\bx_j|},\quad i\ne j\\
{\bf W}_{ii}&=&-\sum_{j\ne i}{\bf w}_{ij}\\
{\bf W}_{ij}&=&{\bf w}_{ij},\quad i\ne j
\eea
If the approach of the previous section is now applied to this metric
to find time-dependent homothetic solutions then, due to the sign changes,
the upshot is that equation (\ref{geo}) is once again obtained, but
with the replacement $C\mapsto -C.$ In this case the physically acceptable
solution is therefore to choose $C=-1.$ With this choice both the 
function $\alpha(t)$ and the positions $\by_k$ that we have found to
provide
geodesics for the Lee-Weinberg-Yi metric carry over unchanged to produce
geodesics of the Gibbons-Manton metric. The difference now is that these
geodesics are only valid in the region where all the monopole are well
separated, so the solutions break down before they can describe the
collision of the monopoles. 

For example, for $n=4,$ the scaling geodesic
describes the scattering of four monopoles on the vertices of a 
contracting tetrahedron. In fact, by using symmetry arguments, the
full geodesic, to which this is a good approximation in the well separated
regime, has been found and shows that as the monopoles approach they
pass through a monopole solution with cubic symmetry and emerge on
the vertices of an expanding tetrahedron which is dual to the
incoming one \cite{HS1}. For this example, even the metric is 
known exactly in terms of elliptic integrals \cite{BS}. 

It might be amusing to attempt to identify the outcomes of the various
$SU(2)$ monopole scatterings that begin as the contracting polyhedra
that we have identified, particularly those with high symmetry. 
However, a note of caution must be applied in this situation.
The Gibbons-Manton metric possesses an $n$-torus isometry which
 the true monopole
metric does not have for any finite separation. This means that the
symmetry of a contracting polyhedron may only be realized in the
true $SU(2)$ monopole solution at the limit of infinite separation.
As an example of this situation consider the case $n=6,$ where the
contracting polyhedron is an octahedron. Using the one-to-one
correspondence \cite{Ja} between $SU(2)$ $n$-monopoles and
 (an equivalence class of) rational maps between Riemann spheres of degree $n$
we may determine the dimension of the moduli space of octahedrally symmetric
$SU(2)$ monopoles of charge six. Degree six polynomials form the carrier space
for the 7-dimensional irreducible representation of $SU(2)$ and when
this representation is restricted to the octahedral group it decomposes
into irreducible representations of the octahedral group of dimensions
one, three and three. Since two polynomials are required to form a 
rational map this shows that there are no octahedrally symmetric rational
maps of degree six, and hence no octahedrally symmetric charge six monopoles.
Thus six $SU(2)$ monopoles placed on the vertices of an octahedron break
the octahedral symmetry for any finite value of the separation, no
matter how large. This contrasts with the above mentioned case of
$n=4$ with tetrahedral symmetry. The 5-dimensional irreducible representation
of $SU(2)$ when restricted to the tetrahedral group decomposes into
two 1-dimensional representations and a 3-dimensional representation.
The basis polynomials for the two 1-dimensional representations yield
a 1-parameter family of tetrahedrally symmetric degree three rational maps,
which corresponds to the geodesic describing the tetrahedral scattering
of four monopoles \cite{HMS}. Thus, in some cases the symmetry of the
scaling geodesics of the Gibbons-Manton metric may be true symmetries of
related geodesics in the true moduli space and in others they may not.

\section{Conclusion}\news
The scattering of $n$ distinct fundamental monopoles can be
approximated by geodesic motion on the Lee-Weinberg-Yi manifold.
We have described a variational method to construct some geodesics on
 this manifold for arbitrary values of $n,$ and applied it to obtain a
number of examples. The energy function used in this approach has
features similar to that which arises in the  classic problem of
 determining central 
configurations, but in contrast to central configurations
 it yields 
points which lie on a single shell for arbitrary values of $n.$
The geodesics constructed by our method describe the scattering of
 monopoles on the
vertices of a contracting, and then expanding, polyhedron, which generically
is a deltahedron. We have found, and exploited, 
similarities between the deltahedra
obtained here and those which arise in the problem of maximizing
the sum of the mutual separations for points on a sphere. 

\section*{Acknowledgements}
We thank Ed Saff and Nikolay Andreev for useful correspondence and
acknowledge advanced fellowships from PPARC (RAB) and EPSRC (PMS).\\

\end{document}